\documentclass{article}

\usepackage{arxiv}

\usepackage[utf8]{inputenc} 
\usepackage[T1]{fontenc}    
\usepackage{hyperref}       
\usepackage{url}            
\usepackage{booktabs}       
\usepackage{amsfonts}       
\usepackage{nicefrac}       
\usepackage{microtype}      
\usepackage{lipsum}
\usepackage{graphicx}
\graphicspath{ {./images/} }

\title{A review of transcranial magnetic stimulation and Alzheimer's disease}

\author{
 Arsalan Heidarpanah \\
  Department of Biomedical Engineering\\
  Islamic Azad University, South Tehran Branch\\
  Tehran, Iran \\
  \texttt{st\_a\_heidarpanah@azad.ac.ir} \\
}

\begin{document}
\maketitle
\begin{abstract}
Since four decades ago, that the transcranial magnetic stimulation (TMS) technique was introduced, the increasing attention of neuroscience researchers and medical engineers has been focused on the development of this technique and its use to manage the treatment of a wide range of neurological conditions, including Alzheimer's disease. The ability of TMS, specifically a substantial type known as repetitive transcranial magnetic stimulation (rTMS) in changing the plasticity of the cortex, has been the most important feature that has created the hopes of controlling Alzheimer's disease with this technique more than ever. 
\end{abstract}


\section{Introduction}
Repetitive transcranial magnetic stimulation (rTMS) has emerged as a powerful tool for non-invasively inducing plasticity in the human cortex and has opened a new perspective for its therapeutic use in a wide range of neurological disorders.
Despite its growing popularity in basic and clinical research, we still know little about how rTMS affects the human brain. Most of the researches use the amplitude changes of motor evoked potentials (MEPs), which are used by single-pulse transcranial magnetic stimulation as a neurophysiological index of neuroplasticity induction after rTMS.

TMS is a non-invasive method to stimulate the brain by passing a temporary magnetic field through the skull and causes an electric current in the underlying cortical tissue.
There is evidence that TMS techniques, specially rTMS, can be used to modulate the activity of the human brain. It has been almost three decades since the early studies of rTMS as a non-invasive method in patients with focal epilepsy [7] .
Other studies have shown that rTMS can be used to improve motor performance in patients with Parkinson's disease [9] . Furthermore, when rTMS apply to the prefrontal cortex (DLPFC) for several days, it can be a beneficial treatment option that has fewer adverse effects than electrical stimulation treatment for patients with drug-resistant depression.

Several studies used MEPs measured from rTMS to explore the electrophysiological effects of different stimulation parameters. It appears that rTMS tends to cause a change in cerebral cortex excitability that decreases with low-frequency rTMS (<1 Hz) and increases with high-frequency rTMS (> 10 Hz).
Recently, in an effort to use stimulation protocols to induce synaptic plasticity in animal models [14], various studies have conducted using high-frequency repetitive stimulation, known as theta burst stimulation (TBS).
Huang et al. (15) adapted the TBS technique for use in humans, and reported that MEP amplitudes increased during sequential stimulation (eg, continuous theta stimulation, cTBS).

The expectation that facilitatory and inhibitory protocols can restore normal brain function by increasing and decreasing cortical excitability, respectively, has created a basis for the use of rTMS in various neurological and psychological conditions [1.] Using neuroimaging approaches like fMRI and EEG to measure rTMS-induced neuroplasticity has advanced our understanding of the mechanisms of rTMS effects on the human brain; It also promises us to provide new ways to target the therapeutic use of rTMS in clinical disorders.
The problem of individual response variability in rTMS research has received considerable attention in the past decade, with a number of MEP studies failing to replicate the results of the previous studies.
In many ways, it should not be surprising that the changes in neural activity caused by rTMS are highly variable. Compared to the protocols used to induce long-term potentiation (LTP) and long-term depression (LTD) in animal models, rTMS applied through the human scalp is much more precise and activates a diffuse cortical network of different cell types. The evidence from various experiments shows that the boundary between LTP and LTD is not clear.
Similarly, it is now well recognized that the level of the effects of rTMS is susceptible to many experimental and biological factors, that not only depends on inter-individual variability (e.g., age, gender, genetics, physical activity levels, etc.) but also it depends on session-related factors (e.g., duration of session, history of synaptic activity, etc.) [20].

It is worth considering an alternative possibility: whether rTMS has no lasting effect on human neural activity, and whether our response variability is only a random noise.
While we do not believe that this is true for all rTMS protocols, it cannot be simply dismissed with the available evidence. Further research involving larger sample sizes, sham control conditions, and repeated assessments in the same subject under the same test conditions to maximize transparency as well as minimize bias [21], can determine the type of rTMS required. Finally, in regard to conducted studies, it appears rTMS significantly alter neural activity in humans.

Nevertheless, while the variable effects of rTMS are a significant challenge for the field, there are still good reasons to be optimistic about its future in basic and clinical research.
Alzheimer's disease (AD) is one of the most common neurological disorders, which is characterized by a decrease in cognition, and interruption of daily routine activities.
AD severity can range from a very early preclinical stage to late-stage dementia with multi-domain cognitive and functional impairments along a biological and clinical continuum.
The syndrome of mild cognitive impairment (MCI) occurs between these stages and represents a landmark from the stage without symptoms to the onset of dementia [22]. Among several clinical subtypes of MCI [23], it is assumed that amnestic MCI (aMCI) has a destructive cause and is more likely to turn into AD dementia [24, 23]. Considering the limited effectiveness of existing drugs to restore brain function, AD is known as one of the main foci in the field of non-pharmacological interventions.
Cognitive training (CT) is a non-pharmacological intervention that is usually recommended in AD and is known as an important adjunctive treatment or even an alternative to pharmacological intervention [26, 25] during the late stage [33-27] with promising results. Failure to remember names is one of the first distinguishing symptoms of episodic memory impairment in AD patients.

This issue increases with the progression of the disease throughout the spectrum of AD, from the early stages to dementia [34]. Several neuroimaging studies have shown that the memory of face-name association consists of a complex network.
In addition to playing an essential role in episodic memory, specialized visual areas in the occipitotemporal cortex and other cortical areas related to higher cognitive functions, like the prefrontal cortex (DLPFC) [35] are a central pole for integrating networks, that mediate organizational functions and may function in different types of tasks. Cognitive changes and DLPFC dysfunction are prominent features of AD in its early stages [36-38] DLPFC is a key region that contributes to several large-scale brain networks, like default mode network (DMN), fronto-parietal network (FPN) and central executive network (CEN) 19, 18, that its changes are related to clinical manifestations of AD. Along with cognitive interventions, repetitive transcranial magnetic stimulation (rTMS) is an emerging and promising therapeutic option in the field of non-pharmacological treatments for AD [41-46].

In the last two decades, rTMS has received increasing attention as a potential therapeutic tool for the treatment of several neurological and neuropsychiatric disorders [47]. rTMS is a non-invasive technique capable of stimulating a magnetic pulse through a coil placed on the head. The created magnetic field induces a transient electric field on the subsurface that can depolarize cortical neurons [49, 48]. Interestingly, rTMS not only acts locally on interneuronal circuits, but also induced activation across cortical connections to functional brain regions [50]. In addition, rTMS is able to produce long-term changes in cortical excitability, that may indicate mechanisms similar to long-term potentiation. [51.] This evidence has aroused great interest in the therapeutic use of rTMS in various clinical fields [53,52]. Existing studies of rTMS in AD suffer from several shortcomings including small sample size, changes in stimulation parameters, target areas, number of sessions and outcome measures, heterogeneity of patients' disease severity, and lack of control groups.

Recent evidence-based guidelines have not confirmed left DLPFC rTMS as an effective treatment option for AD treatment, while it shows the possible effect of rTMS combined with CT in improving cognitive functions in AD patients [54].

The rationale for promoting rTMS as an adjunctive therapy relies on the results of several studies showing that the most effective way to increase neural plasticity (i.e., our brain's ability to change functions and structure by modulating synaptic connections) is a combination of exogenous and endogenous stimulation. In this sense, rTMS may be used as a preparation tool that can pre-activate the initial state of the system so that the neural impact of any subsequent intervention depends on the interaction of ongoing brain activity [52]. This mechanism have a central role in cognitive intervention, where adding rTMS to CT protocol may be a key role to enhance its efficacy [53].

Padullés et al. [59] first showed the beneficial effect of high-frequency rTMS on the left frontal cortex using memory-name association among MCI patients. Since then, the DLPFC has been the target of most rTMS interventions in different stages of AD pathology [66-58]. A recent meta-analysis showed the effects of rTMS in the DLPFC in MCI and AD patients, with high-frequency rTMS protocols in the left DLPFC significantly improve memory performance [42] (as well as low-frequency protocols in the right DLPFC). In the present study, our aim is to review the effect of high-frequency rTMS of the left DLPFC in patients with memory deficits.

\section{rTMS Effect on Neural Diversity for Clinical Research}
\label{sec:headings}

Deactivating the lateral magnocellular nucleus of the anterior neostriatum located directly in motor pathways, prevents the normal modulation of changes and reduces learning capacity [68-71].
Similarly, in people, behavioral diversity increases during the initial learning of new tasks, and it is thought that this process is done by the release of dopamine in the cortico-basal ganglia circuit [73,72].

To support this idea, it is notable that Parkinson's disease characterized by a decrease in dopamine, is associated with movement stiffness and impaired learning ability [75,74].
Additional evidence that movement variability can be beneficial for learning in subjects was shown in a study by Wu et al. [76] They found that the structure of movement-to-movement variability in motor output can predict individual differences in motor learning ability in two different tasks, one reward-based and the other error-based.

Similar to variation and vocal learning in songbirds, task-related variation in motor control circuits can facilitate learning in humans. This concept is naturally extended to neurorehabilitation, where, as a result, increasing movement variety during exercise, it can potentially help restore movement functions lost due to injury.
The present results provide evidence that rTMS is used as an adjunctive tool to increase the effects of name-related memory training and generalization to spatial reasoning in Alzheimer's disease. Future studies combining rTMS with CT protocols and focusing on different cognitive domains (e.g., executive functions) are needed to further investigate the beneficial effect of adjunctive rTMS. In addition, the present findings showed that the level of rTMS additive effect depends on the severity of the disease as well as the level of education of the patients. This finding is important for maximizing treatment effects, as patients with different degrees of cognitive impairment may benefit differently from rTMS. Studies on patients in the early stage of AD, as well as the long-lasting follow-up may address the key question of whether rTMS treatments can delay development of the disease or even stop it.

\section{References}

[1] Lefaucheur JP, André-Obadia N, Antal A, Ayache SS, Baeken C, Benninger DH, Cantello RM, Cincotta M, de Carvalho M, De Ridder D, Devanne H. Evidence-based guidelines on the therapeutic use of repetitive transcranial magnetic stimulation (rTMS). Clinical Neurophysiology. 2014 Nov 1;125(11):2150-206.

[2] Lefaucheur JP, Aleman A, Baeken C, Benninger DH, Brunelin J, Di Lazzaro V, Filipović SR, Grefkes C, Hasan A, Hummel FC, Jääskeläinen SK. Evidence-based guidelines on the therapeutic use of repetitive transcranial magnetic stimulation (rTMS): an update (2014–2018). Clinical neurophysiology. 2020 Feb 1;131(2):474-528.

[3] Ridding MC, Rothwell JC. Is there a future for therapeutic use of transcranial magnetic stimulation?. Nature Reviews Neuroscience. 2007 Jul;8(7):559-67.

[4] Hamada M, Murase N, Hasan A, Balaratnam M, Rothwell JC. The role of interneuron networks in driving human motor cortical plasticity. Cerebral cortex. 2013 Jul 1;23(7):1593-605.

[5] Hordacre B, Goldsworthy MR, Vallence AM, Darvishi S, Moezzi B, Hamada M, Rothwell JC, Ridding MC. Variability in neural excitability and plasticity induction in the human cortex: a brain stimulation study. Brain stimulation. 2017 May 1;10(3):588-95.

[6] López-Alonso V, Cheeran B, Río-Rodríguez D, Fernández-del-Olmo M. Inter-individual variability in response to non-invasive brain stimulation paradigms. Brain stimulation. 2014 May 1;7(3):372-80.

[7] Pascual‐Leone A, Gates JR, Dhuna A. Induction of speech arrest and counting errors with rapid‐rate transcranial magnetic stimulation. Neurology. 1991 May 1;41(5):697-702.

[8] Dhuna A, Gates J, Pascual-Leone A. Transcranial magnetic stimulation in patients with epilepsy. Neurology. 1991 Jul 1;41(7):1067-.

[9] Pascual-Leone A, Valls-Sole J, Brasil-Neto JP, Cammarota A, Grafman J, Hallett M. Akinesia in Parkinson's disease. II. Effects of subthreshold repetitive transcranial motor cortex stimulation. Neurology. 1994 May 1;44(5):892-.

[10] George MS, Wassermann EM, Williams WA, Callahan A, Ketter TA, Basser P, Hallett M, Post RM. Daily repetitive transcranial magnetic stimulation (rTMS) improves mood in depression. Neuroreport: An International Journal for the Rapid Communication of Research in Neuroscience. 1995 Oct.				

[11] Pascual-Leone A, Rubio B, Pallardó F, Catalá MD. Rapid-rate transcranial magnetic stimulation of left dorsolateral prefrontal cortex in drug-resistant depression. The Lancet. 1996 Jul 27;348(9022):233-7.

[12] Berardelli A, Inghilleri M, Rothwell JC, Romeo S, Curra A, Gilio F, Modugno N, Manfredi M. Facilitation of muscle evoked responses after repetitive cortical stimulation in man. Experimental brain research. 1998 Sep;122(1):79-84.

[13] Chen RM, Classen J, Gerloff C, Celnik P, Wassermann EM, Hallett M, Cohen LG. Depression of motor cortex excitability by low‐frequency transcranial magnetic stimulation. Neurology. 1997 May 1;48(5):1398-403.

[14] Larson J, Wong D, Lynch G. Patterned stimulation at the theta frequency is optimal for the induction of hippocampal long-term potentiation. Brain research. 1986 Mar 19;368(2):347-50.

[15] Huang YZ, Edwards MJ, Rounis E, Bhatia KP, Rothwell JC. Theta burst stimulation of the human motor cortex. Neuron. 2005 Jan 20;45(2):201-6.

[16] Hirsch JC, Crepel F. Use‐dependent changes in synaptic efficacy in rat prefrontal neurons in vitro. The Journal of physiology. 1990 Aug 1;427(1):31-49.

[17] Liu L, Wong TP, Pozza MF, Lingenhoehl K, Wang Y, Sheng M, Auberson YP, Wang YT. Role of NMDA receptor subtypes in governing the direction of hippocampal synaptic plasticity. Science. 2004 May 14;304(5673):1021-4.

[18] Nishiyama M, Hong K, Mikoshiba K, Poo MM, Kato K. Calcium stores regulate the polarity and input specificity of synaptic modification. Nature. 2000 Nov;408(6812):584-8.

[19] Shen KZ, Zhu ZT, Munhall A, Johnson SW. Synaptic plasticity in rat subthalamic nucleus induced by high‐frequency stimulation. Synapse. 2003 Dec 15;50(4):314-9.

[20] Ridding MC, Ziemann U. Determinants of the induction of cortical plasticity by non‐invasive brain stimulation in healthy subjects. The Journal of physiology. 2010 Jul 1;588(13):2291-304.

[21] Nosek BA, Ebersole CR, DeHaven AC, Mellor DT. The preregistration revolution. Proceedings of the National Academy of Sciences. 2018 Mar 13;115(11):2600-6.

[22] Albert MS, DeKosky ST, Dickson D, Dubois B, Feldman HH, Fox NC, Gamst A, Holtzman DM, Jagust WJ, Petersen RC, Snyder PJ. The diagnosis of mild cognitive impairment due to Alzheimer’s disease: recommendations from the National Institute on Aging-Alzheimer’s Association workgroups on diagnostic guidelines for Alzheimer’s disease. Focus. 2013 Jan;11(1):96-106.

[23] Petersen RC, Stevens JC, Ganguli M, Tangalos EG, Cummings JL, DeKosky ST. Practice parameter: Early detection of dementia: Mild cognitive impairment (an evidence-based review)[RETIRED]: Report of the Quality Standards Subcommittee of the American Academy of Neurology. Neurology. 2001 May 8;56(9):1133-42.

[24] Petersen RC. Mild cognitive impairment as a diagnostic entity. Journal of internal medicine. 2004 Sep;256(3):183-94.

[25] Bahar-Fuchs A, Clare L, Woods B. Cognitive training and cognitive rehabilitation for persons with mild to moderate dementia of the Alzheimer's or vascular type: a review. Alzheimer's research \& therapy. 2013 Aug;5(4):1-4.

[26] Buschert V, Bokde AL, Hampel H. Cognitive intervention in Alzheimer disease. Nature Reviews Neurology. 2010 Sep;6(9):508-17.

[27] Belleville S, Clement F, Mellah S, Gilbert B, Fontaine F, Gauthier S. Training-related brain plasticity in subjects at risk of developing Alzheimer’s disease. Brain. 2011 Jun 1;134(6):1623-34.

[28] Clare L, Wilson BA, Carter G, Roth I, Hodges JR. Relearning face-name associations in early Alzheimer's disease. Neuropsychology. 2002 Oct;16(4):538.

[29] Clare L, Woods B. Cognitive rehabilitation and cognitive training for early‐stage Alzheimer's disease and vascular dementia. Cochrane database of systematic reviews. 2003(4).

[30] Davis RN, Massman PJ, Doody RS. Cognitive intervention in Alzheimer disease: a randomized placebo-controlled study. Alzheimer Disease \& Associated Disorders. 2001 Jan 1;15(1):1-9.

[31] Hampstead BM, Stringer AY, Stilla RF, Deshpande G, Hu X, Moore AB, Sathian K. Activation and effective connectivity changes following explicit-memory training for face–name pairs in patients with mild cognitive impairment: a pilot study. Neurorehabilitation and neural repair. 2011 Mar;25(3):210-22.

[32] Hampstead BM, Stringer AY, Stilla RF, Giddens M, Sathian K. Mnemonic strategy training partially restores hippocampal activity in patients with mild cognitive impairment. Hippocampus. 2012 Aug;22(8):1652-8.

[33] Cotelli M, Manenti R, Brambilla M, Petesi M, Rosini S, Ferrari C, Zanetti O, Miniussi C. Anodal tDCS during face-name associations memory training in Alzheimer's patients. Frontiers in aging neuroscience. 2014 Mar 19;6:38.

[34] Tak SH, Hong SH. Face-name memory in Alzheimer's disease. Geriatric Nursing. 2014 Jul 1;35(4):290-4.

[35] Haxby JV, Hoffman EA, Gobbini MI. The distributed human neural system for face perception. Trends in cognitive sciences. 2000 Jun 1;4(6):223-33.

[36] Braak H, Braak E. Neuropathological stageing of Alzheimer-related changes. Acta neuropathologica. 1991 Sep;82(4):239-59.

[37] Kumar S, Zomorrodi R, Ghazala Z, Goodman MS, Blumberger DM, Cheam A, Fischer C, Daskalakis ZJ, Mulsant BH, Pollock BG, Rajji TK. Extent of dorsolateral prefrontal cortex plasticity and its association with working memory in patients with Alzheimer disease. JAMA psychiatry. 2017 Dec 1;74(12):1266-74.

[38] Kaufman LD, Pratt J, Levine B, Black SE. Executive deficits detected in mild Alzheimer's disease using the antisaccade task. Brain and behavior. 2012 Jan;2(1):15-21.

[39] Agosta F, Pievani M, Geroldi C, Copetti M, Frisoni GB, Filippi M. Resting state fMRI in Alzheimer's disease: beyond the default mode network. Neurobiology of aging. 2012 Aug 1;33(8):1564-78.

[40] Opitz A, Fox MD, Craddock RC, Colcombe S, Milham MP. An integrated framework for targeting functional networks via transcranial magnetic stimulation. Neuroimage. 2016 Feb 15;127:86-96.

[41] Birba A, Ibáñez A, Sedeño L, Ferrari J, García AM, Zimerman M. Non-invasive brain stimulation: a new strategy in mild cognitive impairment?. Frontiers in Aging Neuroscience. 2017 Feb 13;9:16.

[42] Chou YH, That VT, Sundman M. A systematic review and meta-analysis of rTMS effects on cognitive enhancement in mild cognitive impairment and Alzheimer's disease. Neurobiology of aging. 2020 Feb 1;86:1-0.

[43] Gonsalvez I, Baror R, Fried P, Santarnecchi E, Pascual-Leone A. Therapeutic noninvasive brain stimulation in Alzheimer’s disease. Curr Alzheimer Res. 2017 Jan 1;14(4):362-76.

[44] Pini L, Manenti R, Cotelli M, Pizzini FB, Frisoni GB, Pievani M. Non-invasive brain stimulation in dementia: a complex network story. Neurodegenerative Diseases. 2018;18(5-6):281-301.

[45] Wang X, Mao Z, Ling Z, Yu X. Repetitive transcranial magnetic stimulation for cognitive impairment in Alzheimer's disease: a meta-analysis of randomized controlled trials. Journal of Neurology. 2020 Mar;267(3):791-801.

[46] Lin Y, Jiang WJ, Shan PY, Lu M, Wang T, Li RH, Zhang N, Ma L. The role of repetitive transcranial magnetic stimulation (rTMS) in the treatment of cognitive impairment in patients with Alzheimer's disease: a systematic review and meta-analysis. Journal of the neurological sciences. 2019 Mar 15;398:184-91.

[47] Lefaucheur JP, André-Obadia N, Antal A, Ayache SS, Baeken C, Benninger DH, Cantello RM, Cincotta M, de Carvalho M, De Ridder D, Devanne H. Evidence-based guidelines on the therapeutic use of repetitive transcranial magnetic stimulation (rTMS). Clinical Neurophysiology. 2014 Nov 1;125(11):2150-206.

[48] Rossi S, Hallett M, Rossini PM, Pascual-Leone A, Safety of TMS Consensus Group. Safety, ethical considerations, and application guidelines for the use of transcranial magnetic stimulation in clinical practice and research. Clinical neurophysiology. 2009 Dec 1;120(12):2008-39.

[49] Rossini PM, Burke D, Chen R, Cohen LG, Daskalakis Z, Di Iorio R, Di Lazzaro V, Ferreri F, Fitzgerald PB, George MS, Hallett M. Non-invasive electrical and magnetic stimulation of the brain, spinal cord, roots and peripheral nerves: Basic principles and procedures for routine clinical and research application. An updated report from an IFCN Committee. Clinical neurophysiology. 2015 Jun 1;126(6):1071-107.

[50] Bortoletto M, Veniero D, Thut G, Miniussi C. The contribution of TMS–EEG coregistration in the exploration of the human cortical connectome. Neuroscience \& Biobehavioral Reviews. 2015 Feb 1;49:114-24.

[51] Hoogendam JM, Ramakers GM, Di Lazzaro V. Physiology of repetitive transcranial magnetic stimulation of the human brain. Brain stimulation. 2010 Apr 1;3(2):95-118.

[52] Miniussi C, Rossini PM. Transcranial magnetic stimulation in cognitive rehabilitation. Neuropsychological Rehabilitation. 2011 Oct 1;21(5):579-601.

[53] Miniussi C, Vallar G. Brain stimulation and behavioural cognitive rehabilitation: a new tool for neurorehabilitation?. Neuropsychological Rehabilitation. 2011 Oct 1;21(5):553-9.

[54] Lefaucheur JP, Aleman A, Baeken C, Benninger DH, Brunelin J, Di Lazzaro V, Filipović SR, Grefkes C, Hasan A, Hummel FC, Jääskeläinen SK. Evidence-based guidelines on the therapeutic use of repetitive transcranial magnetic stimulation (rTMS): an update (2014–2018). Clinical neurophysiology. 2020 Feb 1;131(2):474-528.

[55] Rabey JM, Dobronevsky E, Aichenbaum S, Gonen O, Marton RG, Khaigrekht M. Repetitive transcranial magnetic stimulation combined with cognitive training is a safe and effective modality for the treatment of Alzheimer’s disease: a randomized, double-blind study. Journal of Neural Transmission. 2013 May;120(5):813-9.

[56] Lee J, Choi BH, Oh E, Sohn EH, Lee AY. Treatment of Alzheimer's disease with repetitive transcranial magnetic stimulation combined with cognitive training: a prospective, randomized, double-blind, placebo-controlled study. Journal of clinical neurology. 2016 Jan 1;12(1):57-64.

[57] Sabbagh M, Sadowsky C, Tousi B, Agronin ME, Alva G, Armon C, Bernick C, Keegan AP, Karantzoulis S, Baror E, Ploznik M. Effects of a combined transcranial magnetic stimulation (TMS) and cognitive training intervention in patients with Alzheimer's disease. Alzheimer's \& Dementia. 2019 Dec 24.

[58] Alcalá-Lozano R, Morelos-Santana E, Cortes-Sotres JF, Garza-Villarreal EA, Sosa-Ortiz AL, Gonzalez-Olvera JJ. Similar clinical improvement and maintenance after rTMS at 5 Hz using a simple vs. complex protocol in Alzheimer's disease. Brain stimulation. 2018 May 1;11(3):625-7.

[59] Solé-Padullés C, Bartrés-Faz D, Junqué C, Clemente IC, Molinuevo JL, Bargalló N, Sánchez-Aldeguer J, Bosch B, Falcón C, Valls-Solé J. Repetitive transcranial magnetic stimulation effects on brain function and cognition among elders with memory dysfunction. A randomized sham-controlled study. Cerebral cortex. 2006 Oct 1;16(10):1487-93.

[60] Cotelli M, Calabria M, Manenti R, Rosini S, Zanetti O, Cappa SF, Miniussi C. Improved language performance in Alzheimer disease following brain stimulation. Journal of Neurology, Neurosurgery \& Psychiatry. 2011 Jul 1;82(7):794-7.

[61] Devi G, Voss HU, Levine D, Abrassart D, Heier L, Halper J, Martin L, Lowe S. Open-label, short-term, repetitive transcranial magnetic stimulation in patients with Alzheimer’s disease with functional imaging correlates and literature review. American Journal of Alzheimer's Disease \& Other Dementias®. 2014 May;29(3):248-55.

[62] Marra HL, Myczkowski ML, Memória CM, Arnaut D, Ribeiro PL, Mansur CG, Alberto RL, Bellini BB, da Silva AA, Tortella G, de Andrade DC. Transcranial magnetic stimulation to address mild cognitive impairment in the elderly: a randomized controlled study. Behavioural neurology. 2015.

[63] Rutherford G, Lithgow B, Moussavi Z. Short and long-term effects of rTMS treatment on Alzheimer's disease at different stages: a pilot study. Journal of experimental neuroscience. 2015 Jan;9:JEN-S24004.

[64] Ahmed MA, Darwish ES, Khedr EM, Ali AM. Effects of low versus high frequencies of repetitive transcranial magnetic stimulation on cognitive function and cortical excitability in Alzheimer’s dementia. Journal of neurology. 2012 Jan;259(1):83-92.

[65] Balconi M. Dorsolateral prefrontal cortex, working memory and episodic memory processes: insight through transcranial magnetic stimulation techniques. Neuroscience bulletin. 2013 Jun;29(3):381-9.

[66] Haffen E, Chopard G, Pretalli JB, Magnin E, Nicolier M, Monnin J, Galmiche J, Rumbach L, Pazart L, Sechter D, Vandel P. A case report of daily left prefrontal repetitive transcranial magnetic stimulation (rTMS) as an adjunctive treatment for Alzheimer disease. Brain Stimulation: Basic, Translational, and Clinical Research in Neuromodulation. 2012 Jul 1;5(3):264-6.

[67] Dhawale AK, Smith MA, Ölveczky BP. The role of variability in motor learning. Annual review of neuroscience. 2017 Jul 7;40:479.

[68] Charlesworth JD, Warren TL, Brainard MS. Covert skill learning in a cortical-basal ganglia circuit. Nature. 2012 Jun;486(7402):251-5.

[69] Kao MH, Doupe AJ, Brainard MS. Contributions of an avian basal ganglia–forebrain circuit to real-time modulation of song. Nature. 2005 Feb;433(7026):638-43.

[70] Ölveczky BP, Andalman AS, Fee MS. Vocal experimentation in the juvenile songbird requires a basal ganglia circuit. PLoS biology. 2005 May;3(5):e153.

[71] Tumer EC, Brainard MS. Performance variability enables adaptive plasticity of ‘crystallized’adult birdsong. Nature. 2007 Dec;450(7173):1240-4.

[72] Costa RM. A selectionist account of de novo action learning. Current opinion in neurobiology. 2011 Aug 1;21(4):579-86.

[73] Costa RM, Lin SC, Sotnikova TD, Cyr M, Gainetdinov RR, Caron MG, Nicolelis MA. Rapid alterations in corticostriatal ensemble coordination during acute dopamine-dependent motor dysfunction. Neuron. 2006 Oct 19;52(2):359-69.

[74] Berardelli A, Rothwell JC, Thompson PD, Hallett M. Pathophysiology of bradykinesia in Parkinson's disease. Brain. 2001 Nov 1;124(11):2131-46.

[75] Mongeon D, Blanchet P, Messier J. Impact of Parkinson's disease and dopaminergic medication on proprioceptive processing. Neuroscience. 2009 Jan 23;158(2):426-40.

[76] Wu HG, Miyamoto YR, Castro LN, Ölveczky BP, Smith MA. Temporal structure of motor variability is dynamically regulated and predicts motor learning ability. Nature neuroscience. 2014 Feb;17(2):312-21.

\end{document}